\documentclass[aps,prl,showpacs,twocolumn,superscriptaddress,floatfix]{revtex4}
\usepackage{graphicx}
\usepackage{amsmath}
\usepackage{amssymb}
\usepackage{bm}
\usepackage{dcolumn}

\begin{document}

\title{Graphite and graphene as perfect spin filters}
\author{V. M. Karpan}
\affiliation{Faculty of Science and Technology and MESA$^+$
Institute for Nanotechnology, University of Twente, P.O. Box 217,
7500 AE Enschede, The Netherlands}
\author{G. Giovannetti}
\affiliation{Faculty of Science and Technology and MESA$^+$ Institute for Nanotechnology,
University of Twente, P.O. Box 217, 7500 AE Enschede, The Netherlands}
\affiliation{Instituut-Lorentz for Theoretical Physics, Universiteit Leiden, P. O. Box
9506, 2300 RA Leiden, The Netherlands}
\author{P. A. Khomyakov}
\affiliation{Faculty of Science and Technology and MESA$^+$ Institute for Nanotechnology,
University of Twente, P.O. Box 217, 7500 AE Enschede, The Netherlands}
\author{M. Talanana}
\affiliation{Faculty of Science and Technology and MESA$^+$ Institute for Nanotechnology,
University of Twente, P.O. Box 217, 7500 AE Enschede, The Netherlands}
\author{A. A. Starikov}
\affiliation{Faculty of Science and Technology and MESA$^+$ Institute for Nanotechnology,
University of Twente, P.O. Box 217, 7500 AE Enschede, The Netherlands}
\author{M. Zwierzycki}
\affiliation{Institute of Molecular Physics, P.A.N., Smoluchowskiego 17, 60-179 Pozna\'n,
Poland.}
\author{J. van den Brink}
\affiliation{Instituut-Lorentz for Theoretical Physics, Universiteit Leiden, P. O. Box
9506, 2300 RA Leiden, The Netherlands}
\affiliation{Institute for Molecules and Materials, Radboud Universiteit Nijmegen, P. O.
Box 9010, 6500 GL Nijmegen, The Netherlands}
\author{G. Brocks}
\affiliation{Faculty of Science and Technology and MESA$^+$ Institute for Nanotechnology,
University of Twente, P.O. Box 217, 7500 AE Enschede, The Netherlands}
\author{P. J. Kelly}
\affiliation{Faculty of Science and Technology and MESA$^+$ Institute for Nanotechnology,
University of Twente, P.O. Box 217, 7500 AE Enschede, The Netherlands}
\date{\today }

\begin{abstract}
Based upon the observations (i) that their in-plane lattice
constants match almost perfectly and (ii) that their electronic
structures overlap in reciprocal space for one spin direction only,
we predict perfect spin filtering for interfaces between graphite
and (111) fcc or (0001) hcp Ni or Co. The spin filtering is quite
insensitive to roughness and disorder. The formation of a chemical
bond between graphite and the open $d$-shell transition metals that
might complicate or even prevent spin injection into a single
graphene sheet can be simply prevented by dusting Ni or Co with
one or a few monolayers of Cu while still preserving the ideal spin
injection property.
\end{abstract}

\pacs{72.25.-b,73.43.Qt,75.47.-m,81.05.Uw,85.75.-d }
%
%
%
%
%

\maketitle

The observation \cite{Baibich:prl88,Binasch:prb89} of giant magnetoresistance
(GMR) in systems where the transmission through interfaces between normal and
ferromagnetic metals (FM) is spin-dependent has driven a major effort to study
spin filtering effects in other systems.
An ideal spin filter would allow all carriers with one spin through but none
with the other spin. Interfaces with half-metallic ferromagnets (HMFs)
\cite{deGroot:prl83} should have this property but progress in exploiting
it has been slow
because of the difficulty of making stoichiometric HMFs with the theoretically
predicted bulk properties and then making devices maintaining these properties
at interfaces \cite{Dowben:jap04}.

If the nonmagnetic metal (NM) is replaced by an insulator (I) or semiconductor
(SC), spin filtering still occurs giving rise to tunneling magnetoresistance
(TMR) in FM$|$I$|$FM magnetic tunnel junctions (MTJs) and spin-injection at
FM$|$SC interfaces.
If the spin-polarization of the ferromagnet is not complete, then the
conductivity mismatch between metals and semiconductors or insulators
has been identified as a serious obstacle to efficient spin injection
\cite{Schmidt:prb00}. It can be overcome if there is a large
spin-dependent interface resistance but this is very sensitive to the
detailed atomic structure and chemical composition of the interface.
Knowledge of the interface structure is a necessary preliminary to
analyzing spin filtering theoretically and progress has been severely
hampered by the difficulty of experimentally characterizing FM$|$I and
FM$|$SC interfaces.

\begin{table}[b]
\begin{ruledtabular}
\caption[Tab1]{
Lattice constants of Co, Ni, Cu, and graphene.
$a_{\rm hex} \equiv a_{\rm fcc} / \sqrt{2} $.
Equilibrium separation ${d_0}$ for a layer of graphene on top of
graphite, Co, Ni or Cu calculated within the local density approximation
(LDA) of density functional theory with $a=2.46$ \AA.
The binding energy $\Delta E(d_0)=E(d=\infty)-E(d_0)$ is the energy
(per interface unit cell) required to remove a single graphene layer
from a graphite stack or from a Co, Ni or Cu (111) surface.
$W$ is the workfunction.}
\begin{tabular}{lllll}
                               & Graphene & Co    & Ni      & Cu     \\
\hline
$a_{\rm fcc}^{\rm expt}$ (\AA) &          & 3.544\footnotemark[1]
                                                  & 3.524\footnotemark[1]
                                                            & 3.615\footnotemark[1]   \\
$a_{\rm hex}^{\rm expt}$ (\AA) & 2.46     & 2.506 & 2.492   & 2.556  \\
$a_{\rm hex}^{\rm LDA}$  (\AA) & 2.45     & 2.42  & 2.42    & 2.49   \\
${d_0}$ (\AA)                  & 3.30     & 2.04  & 2.03    & 3.18   \\
$\Delta E(d_0)$ (eV)           & 0.10     & 0.37  & 0.32    & 0.07   \\
$W_{\rm calc}$ (eV)            & 4.6      & 5.4   & 5.5     & 5.2    \\
$W_{\rm expt}$ (eV)
                               & 4.6\footnotemark[2]
                                          & 5.0\footnotemark[3]
                                                  & 5.35\footnotemark[3]
                                                            & 4.98\footnotemark[3]   \\
\end{tabular}
\label{tableone}
\end{ruledtabular}
\footnotetext[1]{Ref.\onlinecite{Ibach:95}}
\footnotetext[2]{Ref.\onlinecite{Oshima:jpcm97}}
\footnotetext[3]{Ref.\onlinecite{Michaelson:jap77}}
\end{table}

The situation improved with the confirmation of large values of TMR in
tunnel barriers based upon crystalline MgO
\cite{Yuasa:natm04,Parkin:natm04} which had been predicted by detailed
electronic structure calculations \cite{Butler:prb01,Mathon:prb01}.
While the record values of TMR - in excess of 500\% at low temperatures
\cite{Yuasa:apl06} - are undoubtedly correlated with the crystallinity
of MgO, the nature of this relationship is not trivial
\cite{Tsymbal:pms07}. The sensitivity of TMR (and spin injection) to
details of the interface structure \cite{Zwierzycki:prb03,Xu:prb06}
make it difficult to close the quantitative gap between theory and
experiment. In view of the reactivity of the open-shell transition
metal (TM) ferromagnets Fe, Co and Ni with typical semiconductors and
insulators, preparing interfaces where disorder does not dominate the
spin filtering properties remains a challenge. With this in mind, we
wish to draw attention to a quite different material system which
should be intrinsically ordered, for which an unambiguous theoretical
prediction of perfect spin filtering can be made in the absence of
disorder, and which is much less sensitive to interface roughness and
alloy disorder than TMR or spin injection.

\begin{figure}[btp]
\includegraphics[scale=0.45]{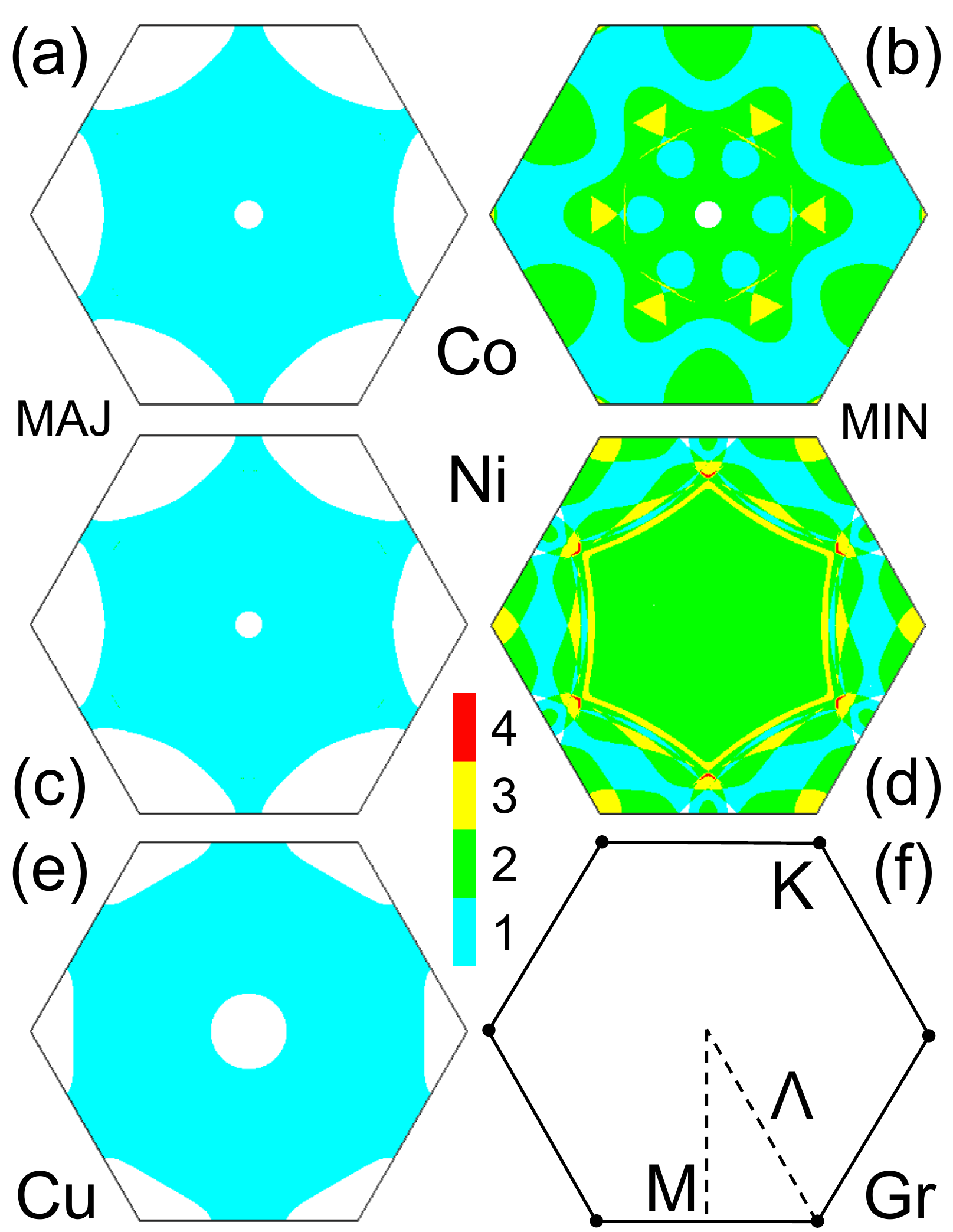}
\caption{Fcc Fermi surface (FS) projections onto a plane perpendicular
to the [111] direction for Co majority (a) and minority (b) spins, for
Ni majority (c) and minority (d) spins and for Cu (e). The number of
FS sheets is shown by the colour bar on the right. For graphene and
graphite, surfaces of constant energy are centred on the K point (f).}
\label{Fig1}
\end{figure}

We begin by observing that the in-plane lattice constants of graphene
and graphite match the surface lattice constants of (111) Co, Ni and
Cu almost perfectly. From Table~\ref{tableone}, it can be seen that Ni
is particularly suitable with a lattice mismatch of only 1.3\%. The
second point to note is that the only electronic states at or close to
the Fermi energy in graphene or graphite are to be found near to the
high symmetry K point in reciprocal space where Co and Ni have states
with minority spin character only. The absence of majority spin states
in a large region about the K point is made clear in the (111) Fermi
surface (FS) projections shown in Fig.~\ref{Fig1}. The (0001) FS
projections for hcp Co are qualitatively the same. It follows that in
the absence of symmetry-lowering (resulting from disorder, interface
reconstruction etc.) perfect spin filtering should occur for graphite
on top of a flat Ni or Co (111) surface.

\begin{figure}[!]
\includegraphics[scale=0.65]{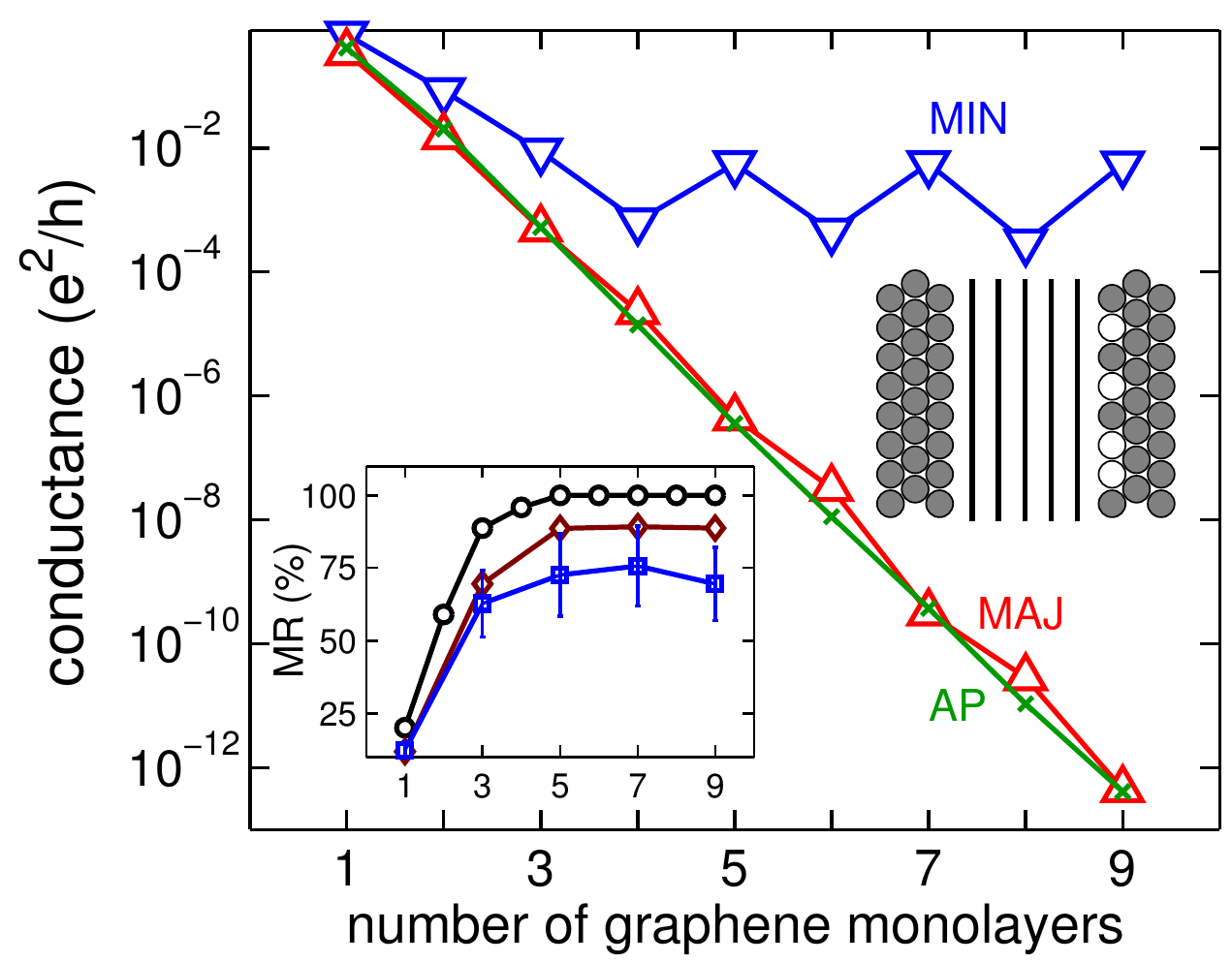}
\caption{Conductances
$G_P^{\rm min}$ ($\triangledown$),
$G_P^{\rm maj}$ ($\vartriangle$), and
$G_{AP}^{\protect\sigma}$ (${\mathbf{\times}}$) of a Ni$|$Gr$_n|$Ni
junction as a function of the number of graphene layers $n$ for ideal
junctions. Inset:
magnetoresistance as a function of $n$ for:
(circles) ideal junctions;
(diamonds) Ni$|$Gr$_n|$Cu$_{50}$Ni$_{50}|$Ni junctions where the
surface layer is a disordered alloy;
(squares) Ni$|$Gr$_n|$Ni junctions where the top layer of one of the
electrodes is rough with only half of the top layer sites occupied
(sketch).
Results for roughness and disorder are modelled in $5 \times 5$ lateral
supercells and averaged over 20 disorder configurations as described in
Refs~\onlinecite{Xia:prb06,Xu:prb06}.
For the rough surface layer, the error bars indicate the spread of MR
obtained for different configurations.
The supercell conductances are normalized to the $1 \times 1$ surface
unit cell used for the ideal case. } \label{Fig2}
\end{figure}

The effectiveness of the spin filtering is tested for a
current-perpendicular-to-the-plane (CPP) structure with $n$ graphene
layers sandwiched between semi-infinite Ni electrodes. The spin
dependent transmission through this Ni$|$Gr$_n|$Ni junction is
calculated using a first-principles tight-binding muffin tin orbital
(TB-MTO) wave-function matching scheme \cite{Xia:prb01,Xia:prb06}
for parallel (P) and antiparallel (AP) orientations of the Ni
magnetizations. The atomic sphere (AS) potentials are calculated
self-consistently within density functional theory for atomic
structures determined by total energy minimization (see below).
The conductances
$G_P^{\sigma}$ and $G_{AP}^{\sigma}$ are shown in Fig.~\ref{Fig2} for the
minority and majority spin channels, $\sigma=$ min, maj.
$G_P^{\rm maj}$ and
$G_{AP}^{\sigma}$ are strongly attenuated while, apart from an
even-odd oscillation,
$G_P^{\rm min}$ is independent of $n$.
The magnetoresistance
MR $=(R_{AP}-R_{P})/R_{AP} \equiv (G_{P}-G_{AP})/G_{P}$ rapidly
approaches 100\%; see inset. We use the \emph{pessimistic}
definition of MR because $G_{AP}$ vanishes for large $n$; usually the
optimistic version is quoted
\cite{Yuasa:natm04,Parkin:natm04,Butler:prb01,Mathon:prb01,Yuasa:apl06}.
Similar results are obtained for Ni$|$Gr$_n|$Co and Co$|$Gr$_n|$Co
junctions.

We envisage a procedure in which thin graphite layers are prepared by
micromechanical cleavage of bulk graphite onto a SiO$_2$ covered Si
wafer \cite{Geim:natm07} into which TM (Ni or Co) electrodes have been
embedded and layers of graphene are peeled away until the desired value
of $n$ is reached. Assuming it will be possible to realize one
essentially perfect interface in a CPP geometry, we studied the effect
of roughness and disorder at the other interface on MR
(inset Fig.~\ref{Fig2}).
Replacing the top Ni layer with a Ni$_{50}$Cu$_{50}$ random alloy only
reduces the MR to 90\% (900\% in the optimistic definition). Extreme
roughness, whereby half of the Ni interface layer is removed at random,
only reduces the MR to 70\%. The momentum transfer induced by the
scattering is apparently insufficient to bridge the large gap about the
K point in the majority spin FS projections. Alternatively, it may be
possible to prepare two separate, near-perfect TM$|$Gr interfaces and
join them using a method analogous to vacuum bonding \cite{Monsma:sc98}.

Graphite has a large $c$-axis resistivity \cite{Matsubara:prb90}. If one
of the TM$|$Gr interfaces is ideal and the graphite layer is sufficiently
thick, then it should be possible to achieve 100\% spin accumulation in
a high resistivity material making it suitable for injecting spins into
semiconductors \cite{Schmidt:prb00}. Because carbon is so light, spin-flip
scattering arising from spin-orbit interaction should be negligible.

The results shown in Fig.~\ref{Fig2}
were calculated for the lowest energy ``AC'' configuration of graphene
on Ni corresponding to one carbon atom above a Ni atom (the surface
``A'' sites) while the other is above a third layer Ni ``C'' site.
A and C refer to the conventional ABC stacking of the layers in an fcc
crystal (AB for an hcp structure). The CPP spin filtering should not
depend on the details of how graphite bonds to the metal surface as
long as the translational symmetry parallel to the interfaces is
preserved. This is confirmed by explicit calculation for the ``AB''
and ``BC'' bonding configurations for varying graphene-metal surface
separation $d$.

\begin{figure}[t]
\includegraphics[scale=0.32]{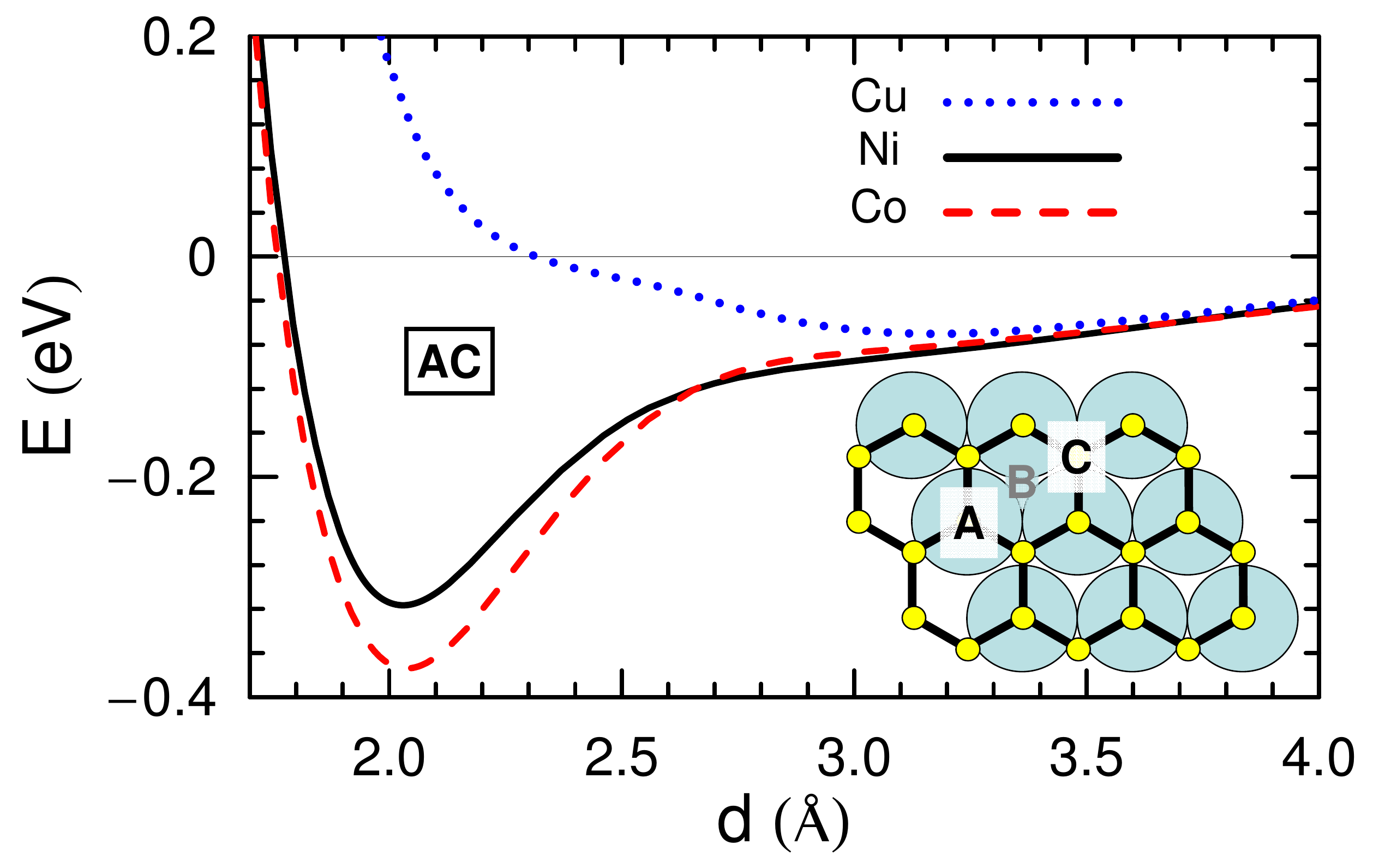}
\caption{Total energy $E$ of a graphene sheet on fcc Co, Ni and Cu
(111) surfaces as a function of the separation of the graphene sheet
from the top layer of the metal. The lowest energy ``AC'' configuration
is sketched on the bottom right.}
\label{E_b}
\end{figure}

\begin{figure}[t]
\includegraphics[scale=0.50]{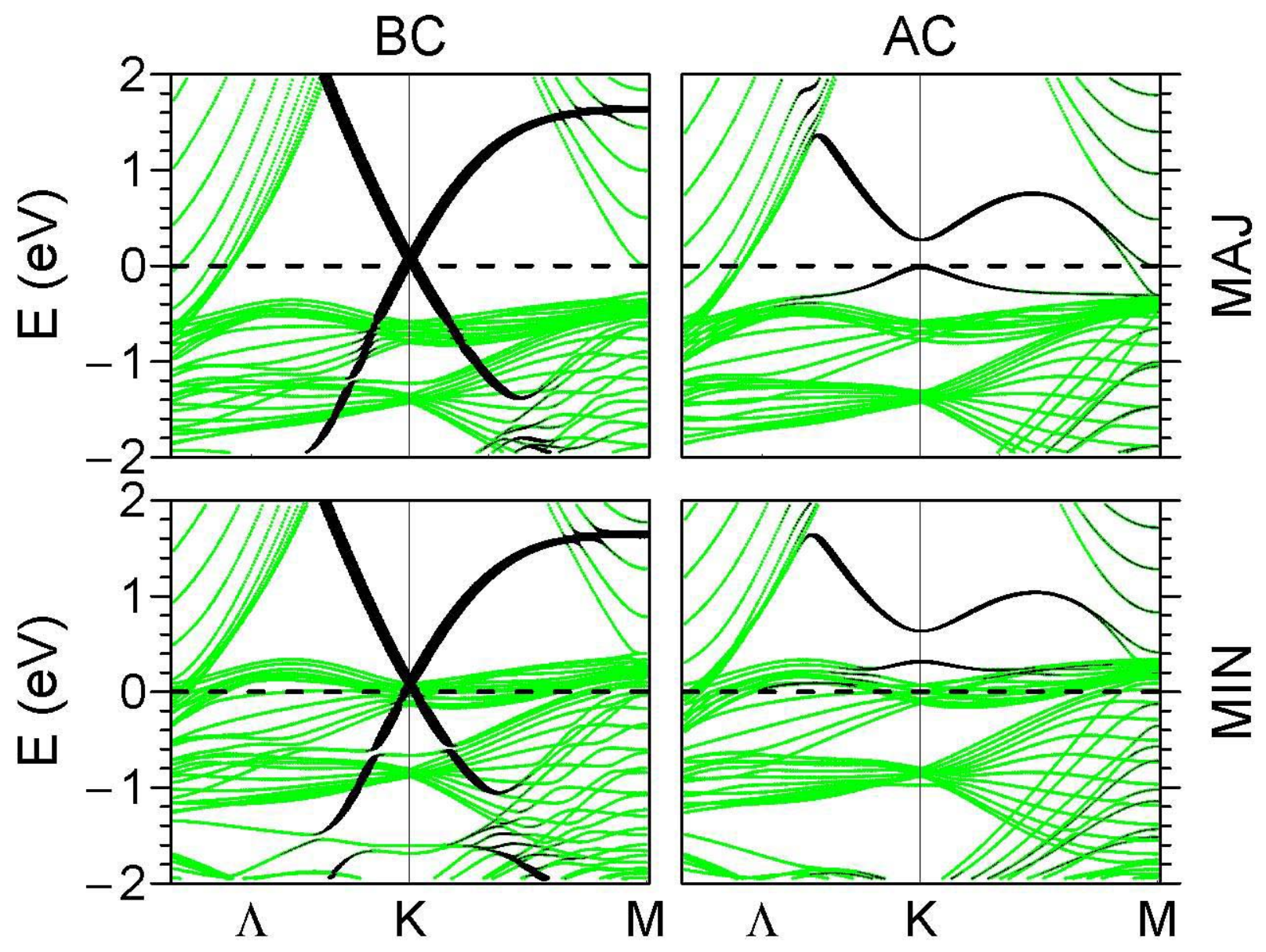}
\caption{Majority and minority spin band structures (green) of a
single graphene layer absorbed upon (both sides of) a 13 layer (111) Ni
slab for a BC configuration with $d=3.3$ \AA, and an AC configuration
with $d=2.0$ \AA. The bands replotted in black using the carbon $p_z$
character as a weighting factor are superimposed.
The Fermi energy is indicated by the horizontal dashed line.}
\label{Bands}
\end{figure}

The nature of the bonding may well play an important role in realizing
such an interface experimentally. In Fig.~\ref{E_b} we show the total
energy of a graphene sheet on TM = Co, Ni and Cu (111) surfaces as a
function of $d$ where the zero of energy has been chosen so that
$E(d=\infty)=0$ for an uncharged graphene sheet.
The density functional theory (DFT) calculations were carried out using
the projector augmented wave (PAW) method \cite{Blochl:prb94,Kresse:prb99},
a plane wave basis set and the LDA, as implemented in the VASP program
\cite{Kresse:prb93,Kresse:prb96}. Supercells containing a slab of at
least six layers of metal atoms with a graphene sheet adsorbed on
one side of the slab and a vacuum region of $\sim 12$ \AA\ were
used. The Brillouin zone of the $(1 \times 1)$ surface unit cell was
sampled using a $36 \times 36$ $\mathbf{k}$-point grid. The plane
wave kinetic energy cutoff was 400 eV. To avoid interactions between
periodic images of the slab a dipole correction was applied
\cite{Neugebauer:prb92}.
The atoms in the metal layers were fixed at their bulk positions. The
experimental lattice constant of graphene $a=2.46$ \AA\ is used as the
lattice parameter $a_{\rm hex}$ for Co, Ni and Cu. To plot the band
structures (Fig.~\ref{Bands}), a 13 layer slab with graphene absorbed
on both sides was used.

The most prominent feature of Fig.~\ref{E_b} is the prediction of a
weak minimum in the binding energy curve for Cu of about 0.07 eV at an
equilibrium separation $d \sim 3.2$ \AA\ and deeper minima of 0.37 and
0.32 eV respectively for Co and Ni, at a smaller equilibrium separation
of $d \sim 2.0$ \AA. In agreement with a recent first-principles
calculation \cite{Bertoni:prb05} and experiment
\cite{Gamo:ss97,Oshima:jpcm97} for graphene on Ni, we find that the
lowest energy corresponds to an AC configuration. The finer details of
the total energy surfaces depend on the choice of exchange-correlation
potential, relaxation of the metal substrate, choice of in-plane lattice
constant etc. and will be presented elsewhere. We restrict ourselves
here to properties which do not depend on these details.

The electronic structure of a single
graphene layer does depend on $d$. For the less strongly bound BC
configuration of Gr on Ni, the equilibrium separation is $d\sim 3.3$
\AA\ and the characteristic band structure of an isolated graphene
sheet is clearly recognizable; see Fig.~\ref{Bands}. For the lowest
energy AC configuration, the interaction between the graphene sheet and
Ni surface is much stronger, a gap is opened in the graphene derived
$p_z$ bands and there are no graphene states at the K point in
reciprocal space at the Fermi energy for the minority spin channel.
This may prevent efficient spin injection into graphene in lateral,
current-in-plane (CIP) devices \cite{Hill:ieeem06}.
However, there is a simple remedy. If a monolayer (or several layers)
of Cu is deposited on Ni, graphene will form only a weak bond with Cu
and the Fermi energy graphene states at the K point will only be weakly
perturbed. Cu will attenuate the conductance of both spin channels
because Cu has no states at the K point but will not change the spin
injection properties as long as it is sufficiently ordered as to
preserve the translational symmetry; a completely mixed CuNi top layer
reduces the MR in a Ni$|$Gr$_n|$Cu$_{50}$Ni$_{50}|$Ni junction only
slightly (Fig.~\ref{Fig2}). Cu will also oxidize less readily than the
more reactive Ni or Co. Spin-flip scattering in a thin layer of Cu
can be neglected. The weaker bonding of graphene to Cu may also have
practical advantages in sample preparation.

Finally, we remark that graphene may exhibit curious bonding properties
to a (111) surface of a Cu$_{1-x}$Ni$_x$ or Cu$_{1-x}$Co$_x$ alloy; as
a function of increasing concentration $x$, the weak minimum at
$d \sim 3.2$ \AA\ will evolve into a deeper minimum at $d \sim 2.0$ \AA\
with the possibility of a double minimum occuring for some range of
concentration $x$; a propensity to form a second minimum is already
evident in the binding energy curve for Cu. Calculations are underway to
examine this possibility.

Planar interfaces between graphene and close-packed Co, Ni, or Cu represent
a very flexible system for studying the influence of atomic and electronic
structure on electrical contact with graphene related systems such as carbon
nanotubes \cite{Schonenberger:sst06} where the nanotube geometry is very
difficult to model using materials specific calculations \cite{Nemec:prl06}.
The binding energy curves in Fig.~\ref{E_b} and electronic structures in
Fig.~\ref{Bands} show that the closer proximity resulting from stronger
bonding does not necessarily lead to better electrical contact if bonding
removes the carbon-related conducting states from the Fermi energy.

Motivated by the recent progress in preparing and manipulating discrete,
essentially atomically perfect graphene layers, we have used
parameter-free, materials specific electronic structure calculations
to explore the bonding and spin transport properties of a novel
TM$|$Gr$_n$ system. We predict perfect spin filtering for ideal
TM$|$Gr$_n|$TM junctions with TM = Co or Ni.

\emph{Acknowledgments:} This work is supported by ``NanoNed'', a
nanotechnology programme of the Dutch Ministry of Economic Affairs. It
is part of the research programs of ``Chemische Wetenschappen'' (CW) and
``Stichting voor Fundamenteel Onderzoek der Materie'' (FOM) and the use of
supercomputer facilities was sponsored by the ``Stichting Nationale Computer
Faciliteiten'' (NCF), all financially supported by the ``Nederlandse
Organisatie voor Wetenschappelijk Onderzoek'' (NWO).
MZ wishes to acknowledge support from EU grant CARDEQ under contract
IST-021285-2.


\end{document}